\documentclass[superscriptaddress,aps,prl,twocolumn]{revtex4}
\usepackage{bm}
\usepackage{ulem}
\usepackage{epsfig}
\usepackage{graphicx}
\usepackage{amssymb,amsmath,amsbsy,amsgen,amsfonts}
\usepackage{dcolumn}
\usepackage{amsthm}
\usepackage{mathrsfs}
\usepackage{latexsym}
\usepackage{array}
\usepackage{color}
\usepackage{amstext}
\allowdisplaybreaks[1]
\usepackage{txfonts}
\usepackage{pifont}
\usepackage{hyperref}

\usepackage{epstopdf} 

\newcommand{\bra}[1]{\left\langle{#1}\right\vert}
\newcommand{\ket}[1]{\left\vert{#1}\right\rangle}

\newcommand{\be}{\begin{equation}}
\newcommand{\ee}{\end{equation}}
\newcommand{\ba}{\begin{array}}
\newcommand{\ea}{\end{array}}
\newcommand{\bqa}{\begin{eqnarray}}
\newcommand{\eqa}{\end{eqnarray}}

\DeclareSymbolFont{symbols}{OMS}{cmsy}{m}{n}

\begin{document}

\title{Experimental characterization of a non-local convertor for quantum photonic networks}

\author{Michal Mi\v{c}uda}
\affiliation{Department of Optics, Palack\'{y} University, 17. listopadu 1192/12,  771~46 Olomouc, Czech Republic}
\author{Robert St\'{a}rek}
\affiliation{Department of Optics, Palack\'{y} University, 17. listopadu 1192/12,  771~46 Olomouc, Czech Republic}
\author{Petr Marek}
\affiliation{Department of Optics, Palack\'{y} University, 17. listopadu 1192/12,  771~46 Olomouc, Czech Republic}
\author{Martina Mikov\'{a}}
\affiliation{Department of Optics, Palack\'{y} University, 17. listopadu 1192/12,  771~46 Olomouc, Czech Republic}
\author{Ivo Straka}
\affiliation{Department of Optics, Palack\'{y} University, 17. listopadu 1192/12,  771~46 Olomouc, Czech Republic}
\author{Miroslav Je\v{z}ek}
\affiliation{Department of Optics, Palack\'{y} University, 17. listopadu 1192/12,  771~46 Olomouc, Czech Republic}
\author{Toshiyuki Tashima}
\affiliation{University of KwaZulu-Natal, School of Chemistry and Physics, Durban 4001, South Africa}
\author{\c{ S}ahin K. \"Ozdemir}
\affiliation{Department of Electrical and Systems Engineering, Washington University in St. Louis, St. Louis, MO 63130 USA}
\author{Mark Tame}
\affiliation{University of KwaZulu-Natal, School of Chemistry and Physics, Durban 4001, South Africa}

\date{\today}

\begin{abstract}
We experimentally characterize a quantum photonic gate that is capable of converting multiqubit entangled states while acting only on two qubits. It is an important tool in large quantum networks, where it can be used for re-wiring of multipartite entangled states or for generating various entangled states required for specific tasks. The gate can be also used to generate quantum information processing resources, such as entanglement and discord. In our experimental demonstration, we converted a linear four qubit cluster state into different entangled states, including GHZ and Dicke states. The high quality of the experimental results show that the gate has the potential of being a flexible component in distributed quantum photonic networks.
\end{abstract}


\maketitle

{\it Introduction.---}
Quantum networks consisting of multipartite entangled states shared between many nodes provide a setting for a wide variety of quantum computing and quantum communication tasks~\cite{Hor,Gis,Gisin07,multi,Chir,Per}. Recent works have experimentally realized some of the basic features of distributed quantum computation~\cite{Barz12,Barz13,Greganti16} and quantum communication schemes, including quantum secret sharing~\cite{Tittel01,Bell14}, open-destination teleportation~\cite{Zhao} and multiparty quantum key distribution~\cite{Chen05,Adamson06}. These experiments employed networks of small-sized entangled resources and showed the potential of distributed quantum information processing in realistic scenarios~\cite{Kimble}. Individual photons serve as a viable platform for implementation of quantum networks, since they can be easily transmitted over free-space or fiber links in order to distribute the necessary resources \cite{multi,OBrien05}. 

A common problem in quantum networks is that once the entangled resource is shared among the nodes it is fixed and can only be used for a given set of quantum tasks~\cite{Kimble,Chir,Per}. A different task then requires conversion of the available multipartite entangled state into another state. When the separation between the nodes is large, such conversion can employ only local operations and classical communication, which severely limits the class of potentially available states~\cite{Dur00,Acin01,Ver02,Kob14}. Fortunately, in some cases two nodes of the network may be close enough for application of a non-local operation between them. This relaxes the constraint and opens an interesting question: what types of entangled states are convertible in this scenario?

Recently, a non-local conversion gate was proposed for exactly this setting of two nodes in close proximity~\cite{Tashima}. It was shown that one can employ a single probabilistic two-qubit gate to convert a four-qubit linear cluster state~\cite{clusterrev} into many other forms of four-qubit entangled states that are inequivalent to each other under local operations and classical communication. The gate therefore enables one to convert between different states so that different tasks can be performed. For instance, the four-qubit linear cluster state can be used for a variety of quantum protocols, such as blind quantum computation~\cite{Barz12} and quantum algorithms~\cite{Walther05,Prevedel07,Tame07,Vallone10,Lee12,Tame14}. On the other hand, a four qubit GHZ state~\cite{GHZ} can be used for open-destination teleportation~\cite{Zhao} and multiparty quantum key distribution~\cite{Adamson06}, and a four-qubit Dicke state can be used for telecloning~\cite{Kiesel07} and quantum secret sharing~\cite{Dicke1,Dicke2}. In this work, we experimentally realize the non-local conversion gate of Ref.~\cite{Tashima} with single photons using a linear optical setup and characterize its performance using quantum process tomography. We find that the conversion gate operates with high quality under realistic conditions and show its potential for converting a four-qubit linear cluster state into a GHZ state, a Dicke state~\cite{Kiesel07}, and a product of two Bell states~\cite{Hor}. The conversion gate can also be used to generate quantum correlations that are not associated with entanglement, but whose presence is captured by the notion of discord~\cite{Mod}. The generated states with discord may also be used as resources in distributed quantum tasks~\cite{Dakic,Gu,Alm,Pirandola}. Furthermore, the conversion gate can be used for `re-wiring' the entanglement connections in a larger graph state network~\cite{clusterrev,Hein}. The experimental results match the theory expectations well and highlight the suitability of the conversion gate as a flexible component in photonic-based quantum networks.

\begin{figure}[t]
\centering
\includegraphics[width=7.5cm]{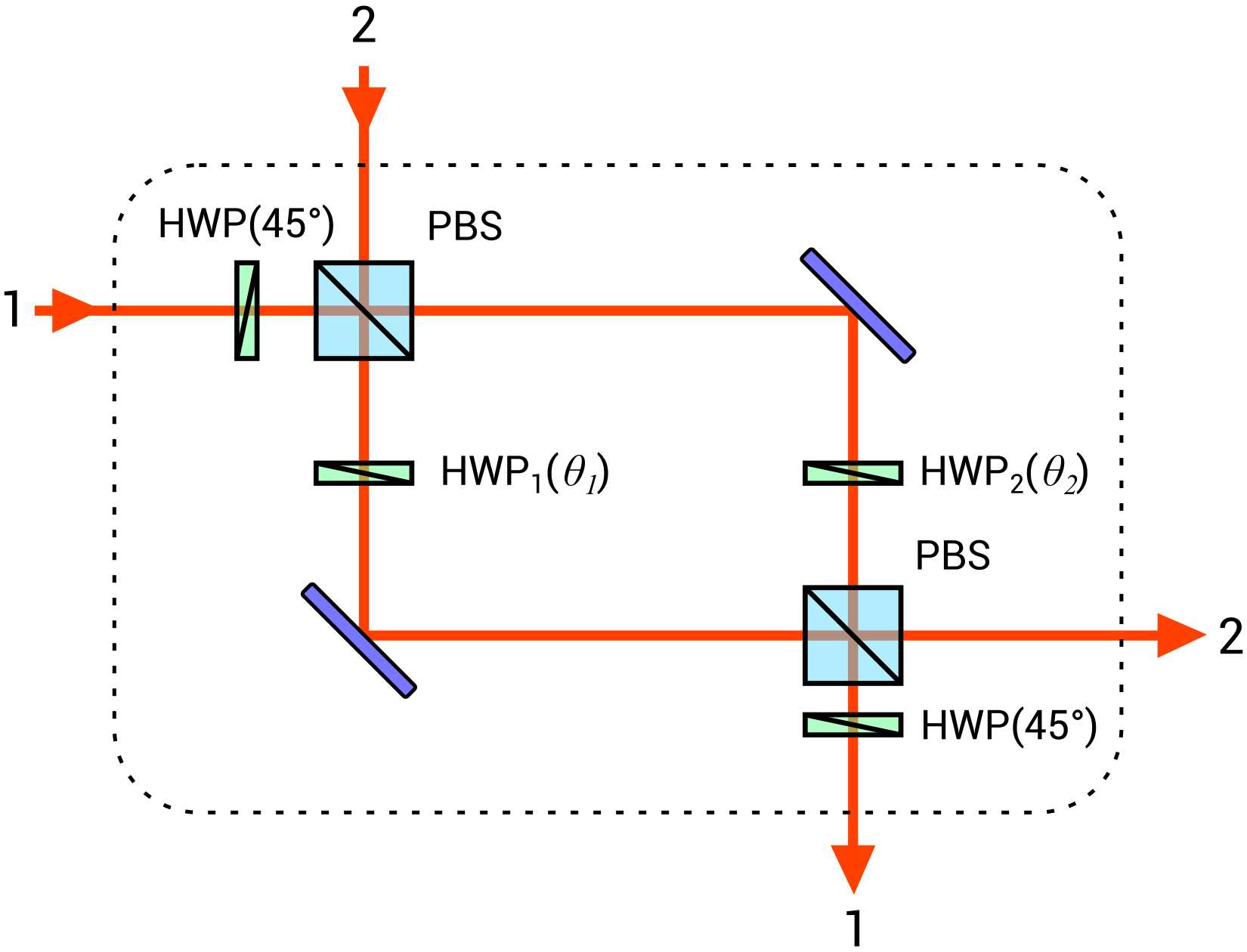}
\caption{The non-local conversion gate using linear optics. The gate takes two photons as inputs, one in each input mode, and performs a non-local operation when the photons exit from different output ports. PBS - polarizing beam splitter, and HWP - half-wave plate.}
\label{fig1}
\end{figure}

{\it Theoretical background.---}
The non-local conversion gate for polarization encoded photonic qubits is depicted in Fig.~\ref{fig1}. The gate operation is based on postselection where one photon is detected at each of the output ports. The gate itself is based on a Mach-Zehnder interferometer and it is created from two polarizing beam splitters (PBSs) and four half-wave plates (HWPs).
Two of the HWPs labeled as HWP($45^{\circ}$) are rotated to a fixed angle of $45^{\circ}$, the other two HWPs labeled as HWP$_1(\theta_1$) and HWP$_2(\theta_2$) are used to adjust the gate to a particular setting. The total operator characterizing the action of the gate in the computational basis of horizontally ($\ket{H}$) and vertically ($\ket{V}$) polarized photons is given by
\bqa\label{Goperator}
\nonumber
\widehat{G}(\theta_1,\theta_2)  &=& (\alpha_1-\beta_1)\ket{HH}_{out \; in} \! \bra{HH} \\ \nonumber
&+&  (\alpha_2-\beta_2) \ket{VV}_{out \; in} \! \bra{VV} \\ \nonumber
&+& \mu_1 \ket{HV}_{out \; in} \!\bra{HV} - \mu_2 \ket{VH}_{out \; in} \!\bra{VH} \\
&+& \mu_1 \ket{VH}_{out \; in} \!\bra{VH} - \mu_2 \ket{HV}_{out \; in} \!\bra{HV},
\eqa
where $\alpha_k = \cos^2(2\theta_k)$ and $\beta_k = \sin^2(2\theta_k)$ ($k=1,2$), and $\mu_1 = \cos(2\theta_1)\cos(2\theta_2)$ and $\mu_2 = \sin(2\theta_1)\sin(2\theta_2)$. The input modes are labelled $in=1,2$ and the output modes are labelled $out=1,2$.

The main feature of the non-local conversion gate is the ability to convert quantum states from one type to another, even though such a conversion is impossible with local operations and classical communication.  To demonstrate the gate's capabilities consider a four-qubit linear cluster state $\ket{C_4}$ given by
\be\label{C4linear}
\ket{C_4} = \frac{1}{2}(\ket{HHHH}+\ket{HHVV} + \ket{VVHH}-\ket{VVVV})
\ee
as an input. Applying the gate on the second and third qubits yields
\bqa\label{C4con}
\nonumber
\widehat{G}(\theta_1,\theta_2) \ket{C_4} &=& \frac{1}{2}[(\alpha_1-\beta_1)\ket{HHHH} - (\alpha_2-\beta_2)\ket{VVVV} \\ \nonumber
& & + \mu_1\ket{HHVV} + \mu_1\ket{VVHH}  \\
& & - \mu_2\ket{HVHV} - \mu_2\ket{VHVH}].
\eqa
The angles $\theta_1$ and $\theta_2$ can be tuned in order to achieve the conversion of the cluster state to a specific kind of state. Notable examples are the four-qubit GHZ and Dicke states, as well as a pair of maximally entangled bipartite states. The angle settings and success probabilities for these example conversions can be found in Tab.~\ref{tableI}.

The non-local nature of the gate can also be effectively utilized to generate classical or nonclassical correlations in a pair of initially separable states \cite{Tashima}. For example, the gate $\widehat{G}(3\pi/8,\pi/8)$ transforms a pair of factorized pure states into a maximally entangled state, while the gate $\widehat{G}(\pi/3,0)$ transforms the state $\frac{1}{2} \openone\otimes |+\rangle \langle +|$, where $|+\rangle=\frac{1}{\sqrt{2}}(|H\rangle + |V\rangle)$,  into a state with quantum correlations but no entanglement.

\begin{table}[t]
\centering
\begin{tabular}{| l | c | c | c |}
\hline \hline
\textbf{Converted state} & $\mathbf{\theta_1}$ & $\mathbf{\theta_2}$ & $p_s$ \\
\hline
Cluster state & 0 & 0 & 1 \\
 & $\pi/2$ & $\pi/2$ & 1 \\
\hline
GHZ state & \, 0  \, & $\pi/4$ & 1/2 \\
 & \,$\pi/2$ \, & $\pi/4$ & 1/2 \\
& $\pi/4$ & \, 0 \, & 1/2\\
& $\pi/4$ & \, $\pi/2$\, & 1/2\\
\hline
Dicke state & $\theta_+$ & $\theta_-$ & \, 3/10 \, \\
 & $\theta_-$ & $\theta_+$ & 3/10 \\
\hline
Two Bell states & $3\pi/8$ & $\pi/8$ & 1/4 \\
 & $\pi/8$ & $3\pi/8$ & 1/4 \\
\hline\hline
\end{tabular}
\caption{Non-local conversion gate angle settings and success probabilities $p_s$ for converting a linear cluster state to GHZ, Dicke and two Bell states. Each kind of conversion can be realized by several settings. The angle $\theta_\pm$ is found from the relation ${\sin(2\theta_\pm)=\sqrt{(5\pm\sqrt{5})/10}}$.}
\label{tableI}
\end{table}

\begin{figure}[t]
\centering
\includegraphics[width=7.8cm]{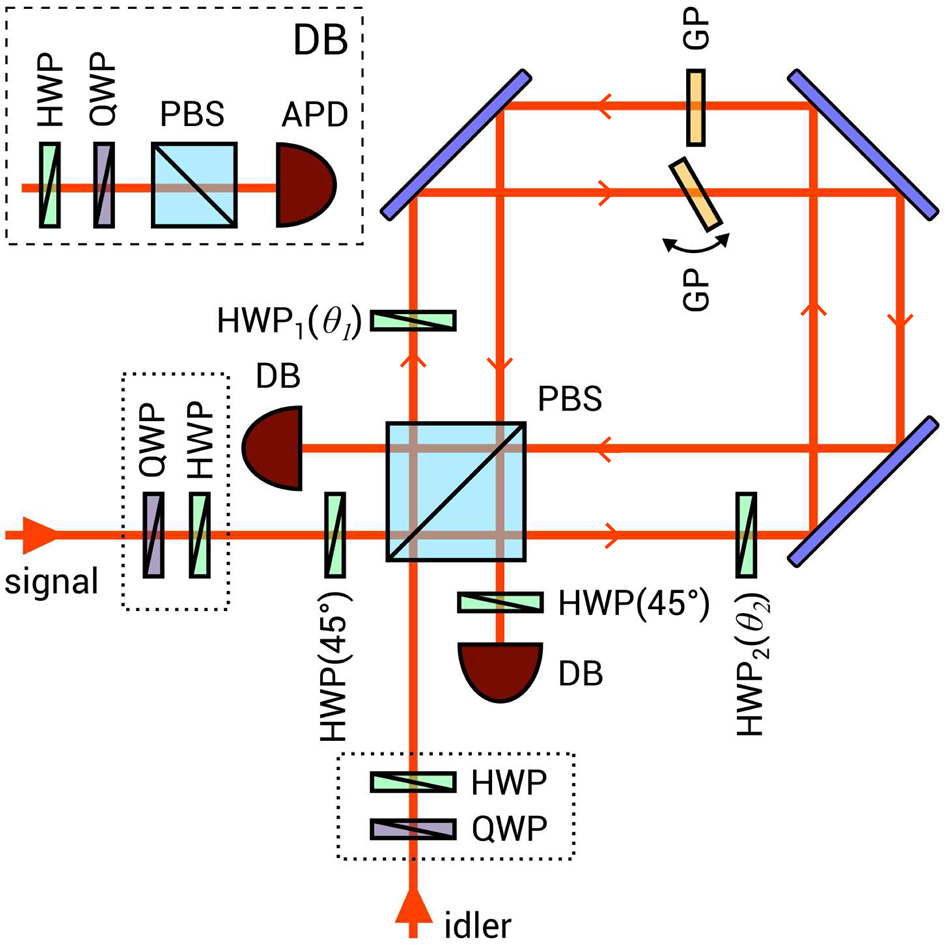}
\caption{Experimental setup for the charaterization of the non-local conversion gate. QWP - quarter-wave plate, HWP - half-wave plate, PBS - polarizing beam splitter, GP - glass plate, DB - detection block, and APD - avalanche photodiode. The dotted boxes represent preparation stages for encoding different inputs. The dashed box represents the analysis stages (DB) for characterizing the output states of the gate.}
\label{fig2}
\end{figure}

{\it Experimental setup.---}
We experimentally demonstrated and characterized the photonic non-local conversion gate using the linear optical setup shown in Fig.~\ref{fig2}. Here, orthogonally polarized time-correlated photon pairs with central wavelength of 810 nm were generated in the process of degenerate spontaneous parametric down-conversion in a BBO crystal pumped by a continuous-wave laser diode~\cite{PDC1,PDC2} and fed into single mode optical fibers guiding photons to the signal and idler input ports of the linear optical setup. The linearly polarized signal and idler photons were decoupled into free space and directed into polarization qubit state preparation blocks (dotted boxes), each consisting of a quarter-wave plate (QWP) and a half-wave plate (HWP). In contrast to the theoretical proposal of Ref.~\cite{Tashima} shown in Fig.~\ref{fig1}, the experimental conversion gate was implemented using a displaced Sagnac interferometer and a single polarizing beam splitter (PBS), where the interferometric phase was controlled by tilting one of the glass plates (GP). This construction provides passive stabilization of the Mach-Zehnder interferometer~\cite{Micuda14}. HWP$_1(\theta_1$) and HWP$_2(\theta_2$) were used to configure the conversion gate for its different settings. Outputs from the conversion gate were analyzed using the detection blocks (DB), which consist of a HWP, a QWP, and a PBS followed by an avalanche photodiode (APD). The scheme operated in the coincidence basis and the operation succeeded upon detecting a two-photon coincidence at the output ports.

By measuring the coincidences we were able to carry out complete quantum process tomography~\cite{Paris04} of the non-local conversion gate for all the settings in Tab.~\ref{tableI}. Each input qubit was prepared in six states $\{|H\rangle,|V\rangle,|+\rangle,|-\rangle,|R\rangle,|L\rangle\}$, and each output qubit was measured in three bases $\{|H\rangle,|V\rangle\}$, $\{|+\rangle,|-\rangle\}$ and $\{|R\rangle,|L\rangle\}$, where $|\pm\rangle=\frac{1}{\sqrt{2}}(|H\rangle \pm |V\rangle)$, $|R\rangle=\frac{1}{\sqrt{2}}(|H\rangle + i |V\rangle)$ and $|L\rangle=\frac{1}{\sqrt{2}}(|H\rangle - i |V\rangle)$. Two-photon coincidences corresponding to the measurement in any chosen product of two-qubit bases were recorded sequentially and the measurement time of each basis was set to 10~s. Using the measured coincidence counts as a mean value of the Poisson distribution from the down-conversion we numerically generated 1000 samples in order to estimate the uncertainty of the experimental results. The process matrices $\chi$ of the quantum process were reconstructed from this data using a Maximum Likelihood estimation algorithm~\cite{Jezek_03,Hradil_04}.

The quality of the different non-local gate operations can be evaluated with help of the process fidelity ${F_{\chi} = \mathrm{Tr}[\sqrt{\sqrt{\chi_{th}}\; \chi \sqrt{\chi_{th}}}]^2}$, which is the overlap between the reconstructed process matrix, $\chi$, and the process matrix for the ideal theoretical operation, $\chi_{th}$. The process matrix $\chi$ represents the completely positive map that fully characterizes the conversion gate operation. Using the Jamiolkowski-Choi isomorphism~\cite{Jam72,Choi75}, the matrix $\chi$ is defined on the tensor product of the input and output Hilbert spaces ${\cal H}_{in}$ and ${\cal H}_{out}$, which are each four-dimensional Hilbert spaces spanning the polarization states of the two photons. Therefore $\chi$ is a $16 \times 16$ matrix and the two-qubit input state $\rho_{in}$ transforms to the two-qubit output state, $\rho_{out}$, according to the relation $\rho_{out}={\rm Tr}_{in}[(\rho_{in}^{T}\otimes \openone_{out})\chi]$, where $T$ denotes transposition. We also use the process purity $P = \mathrm{Tr}[\chi^2]$ to quantify the quality of the operation. For the ideal theoretical case the matrix $\chi_{th}$ corresponds to a pure density matrix, $\chi_{th}=\openone \otimes \hat{G} \ket{\psi^+}\bra{\psi^+} \openone \otimes \hat{G}^\dag$, where $\ket{\psi^+}=\sum_{a,b=H}^V\ket{ab}_{in}\ket{ab}_{out}$ denotes a maximally entangled state on two copies of a two-qubit Hilbert space, and therefore $P=1$.

\begin{figure*}[t]
\centerline{\hspace*{0.0cm}\includegraphics[width=1\linewidth]{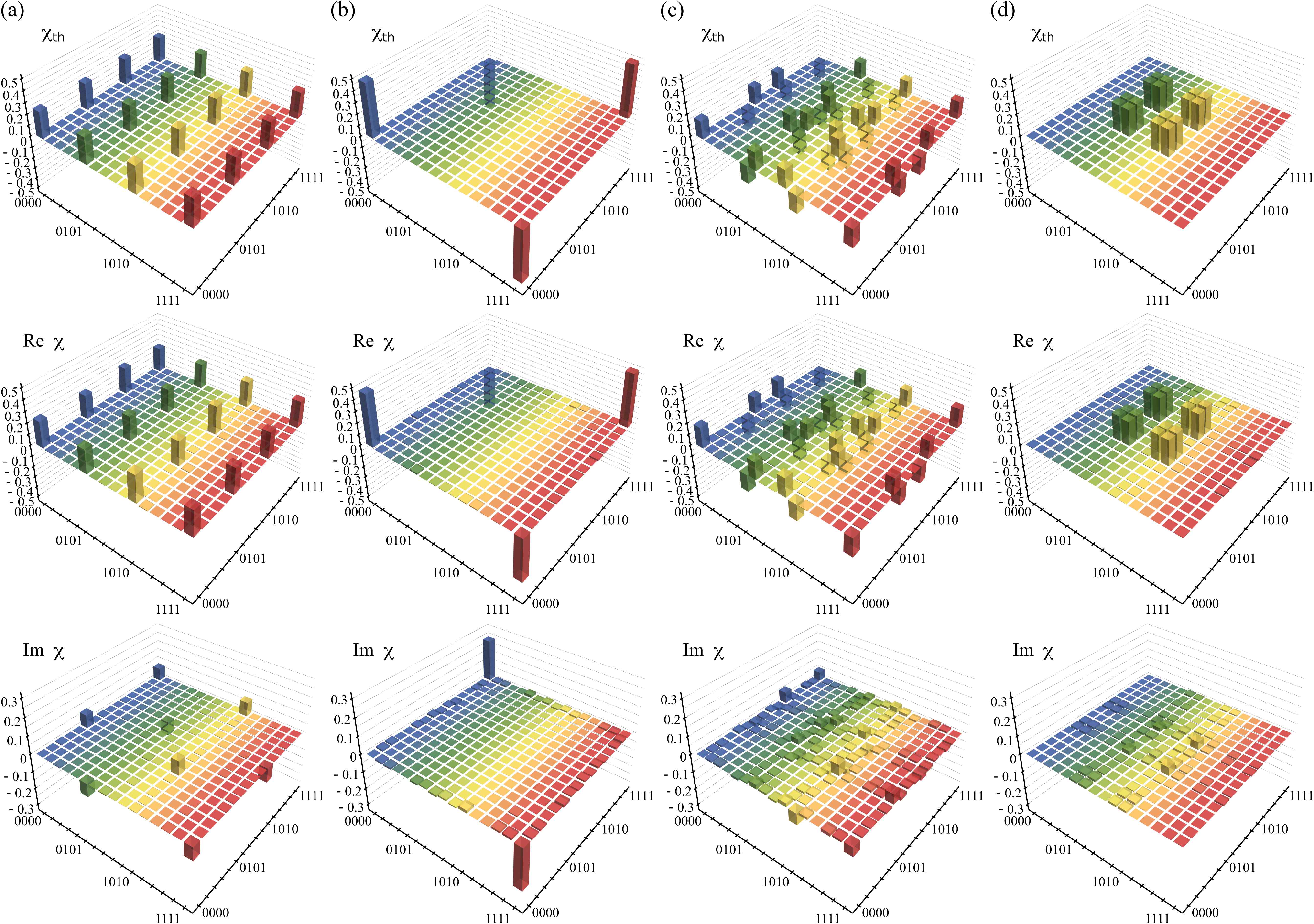}}
\caption{Reconstructed process matrices $\chi$ of the conversion gate using the Jamiolkowski-Choi representation. The $16 \times 16$ matrices are written in the polarization basis of the input and output Hilbert spaces ($\{ \ket{0},\ket{1}\} \leftrightarrow \{ \ket{H},\ket{V}\}$) and correspond to the gate operations given in Tab.~\ref{tableI}. They are arranged in columns: (a) Cluster state, (b) GHZ state, (c) Dicke state, and (d) two Bell states. The theoretical process matrix $\chi_{th}$ is depicted in the top row of each column followed by the real and imaginary part of the reconstructed process matrix in the middle and bottom row, respectively. Note that the theoretical process matrix has only real values.}
\label{fig3}
\end{figure*}

A possible source of reduction in the process fidelity is the introduction of phase shifts experienced by one or more modes in the setup, which are caused by the imperfect nature of the realistic experimental components. These can be compensated for by suitable phase corrections. To reflect this we calculated two kinds of process fidelity for each scenario. The first is the raw fidelity, which was calculated directly from the reconstructed process matrix. The second is the optimized fidelity, which was calculated from the process matrix subject to four phase shifts, one in each of the two input and two output modes. The four phases were optimized over and ultimately chosen in such a way that the resulting fidelity is maximal. The relevant process purity and process fidelity for all the considered scenarios  are given in Tab.~\ref{tableII}, while the process matrices are shown in Fig.~\ref{fig3}.

\begin{table}[t]
\centering
\begin{tabular}{| l | c | c | c |}
\hline \hline
\textbf{Product of } & Purity & Fidelity  & Fidelity \\
\textbf{the conversion} &   &  (raw) & (optimized)  \\
\hline
Cluster state & 0.949(1) & 0.947(1) & 0.973(1)  \\
\hline
GHZ state & 0.913(3) & 0.875(1) & 0.952(2) \\
 \hline
Dicke state & 0.917(2) & 0.925(1) & 0.948(1)  \\
\hline
Two Bell states \, & \, 0.920(2) \, & \, 0.947(1) \, & 0.953(1)  \\
\hline\hline
\end{tabular}
\caption{Process purity and process fidelity of the non-local conversion gate, including one standard deviation related to the last significant digit (represented by the number in brackets).}
\label{tableII}
\end{table}

We also analyzed the conversion gate and its performance from a different angle. The gate is non-local and as such it should be able to transform a two-qubit factorized state into a state with non-zero entanglement. To see this entanglement generation, the input state was set to $|- -\rangle_{\mathrm{in}}$ and fed into the conversion gate with parameters $\theta_1 = 3\pi/8$ and $\theta_2 = \pi/8$, which in the ideal case transforms it into the entangled Bell state $|\Phi^+\rangle=\frac{1}{\sqrt{2}}(\ket{HH}+\ket{VV})$. Using the non-local conversion gate we generated the maximally entangled Bell state with purity $P = 0.946(7)$ and fidelity $F = 0.966(3)$. The number in the brackets represents one standard deviation at the final decimal place.

We can also use the conversion gate to prepare a separable state, {\it i.e.} a state with no entanglement, but with non-zero quantum correlations that can be measured by the discord~\cite{Mod}. For this we started with a mixed factorized state $\rho_{in} = \frac{1}{2} \openone\otimes |+\rangle \! \langle +|$
and fed it into the conversion gate with parameters $\theta_1 =  \pi/3$ and $\theta_2 = 0$. 
The experimental realization was similar to the previous case, only the
totally mixed state was prepared by using an electronically driven fiber
polarization controller. The polarization controller applied mechanical
stress on the input single mode optical fiber in three orthogonal axes
using three co-prime frequencies.
This randomized the polarization state on a time scale of tens of ms, which is two orders of magnitude shorter than the projection-acquisition time of 1~s, thus effectively resulting in a partially mixed state. This preparation method lead to an output state with zero entanglement and non-zero discord. The output state was again determined by using full two-mode quantum state tomography, followed by a maximum likelihood estimation algorithm. Separability of a realistic reconstructed state is difficult to prove, but both entanglement measures we employed - logarithmic negativity  $LN = 0.019(20)$ and concurrence $C = 0.015(15)$ \cite{Hor} - show values separated from zero by less than one standard deviation. This points to a high probability that the state is indeed separable. On the other hand, the discord of the state is $D = 0.066(7)$ \cite{Mod}, which is significantly positive. The confidence intervals were obtained by using a Monte Carlo method based on the measured data.

\begin{figure*}[t]
\centerline{\hspace*{0.0cm}\includegraphics[width=0.8\linewidth]{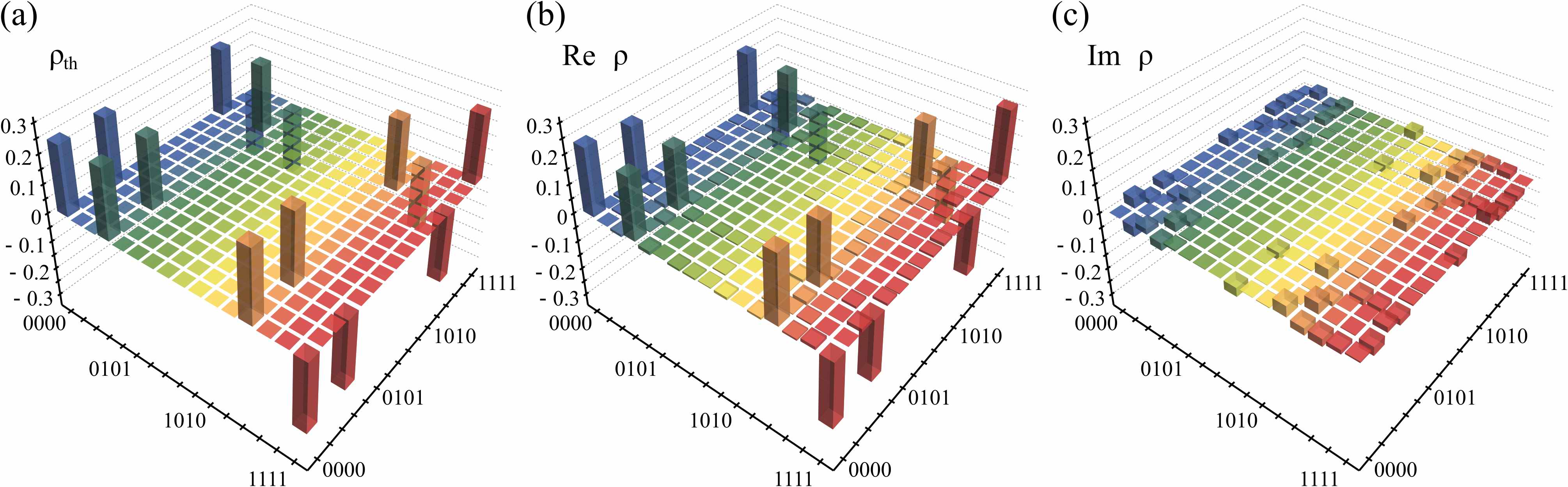}}
\caption{Density matrix of the linear cluster state. (a) Ideal theoretical matrix $\rho_{th}=\ket{C_4}\bra{C_4}$, (b) the real part of the experimental matrix $\rho$, (c) the imaginary part of the experimental matrix $\rho$. Note that the ideal theoretical density matrix has only real values.}
\label{fig4}
\end{figure*}

As the final step of our analysis we looked at how the conversion gate might perform in a realistic scenario. For this, we employed the reconstructed process matrices from Fig.~\ref{fig3} and numerically simulated the effect of the conversion gate on a realistic version of a four-qubit linear cluster state generated in a four-qubit linear-optical quantum logic circuit~\cite{Micuda}. The cluster state $\ket{C_4}\bra{C_4}$, whose density matrix was reconstructed with the help of a maximum likelihood algorithm, is shown in Fig.~\ref{fig4}. The state fidelity, given by ${F = \mathrm{Tr}[\sqrt{\sqrt{\rho_{th}}\; \rho \sqrt{\rho_{th}}}]^2}$, where $\rho_{th}$ is the ideal theoretical state and $\rho$ is the experimental state, gives a value of $F = 0.915$.
\begin{table}[b]
\centering
\begin{tabular}{| l | c | c |}
\hline \hline
\textbf{Product of } &  Fidelity  & \, Fidelity \, \\
\textbf{the conversion} & \, (operation) \, & (total)  \\
\hline
Cluster state & 0.98 & 0.87   \\
\hline
GHZ state & 0.99 & 0.89   \\
 \hline
Dicke state & 0.97 & 0.84   \\
\hline
Two Bell states \, & 0.97 & 0.91  \\
\hline\hline
\end{tabular}
\caption{Fidelity of output states converted from a realistic $\ket{C_4}\bra{C_4}$ state.}
\label{tableIII}
\end{table}
The simulated output state fidelities for the different conversions are given in Tab.~\ref{tableIII}. In the second column, the operation fidelity measures the overlap between the realistic state transformed by the experimental conversion gate and the ideal theoretical conversion gate. In the third column, the total fidelity measures the overall fidelity between the realistic state transformed by the experimental conversion gate and the ideal state transformed by the ideal theoretical conversion gate. The values in Tab.~\ref{tableIII} show the various limits of the gate derived from current experimental technology.

{\it Summary and discussion.---}
We have experimentally realized the non-local photonic conversion gate proposed in Ref.~\cite{Tashima} and tested its performance for all four of its basic conversion settings. We performed quantum process tomography and characterized the individual conversion gate operations by their process matrices. We also directly tested the ability of the conversion gate to generate quantum correlations without entanglement between a pair of separable qubits, finding no entanglement, but non-zero discord present. Finally, we tested the limits of the setup by simulating its action on a realistic experimental four-qubit linear cluster state. In all the tests the conversion gate performed close to the theoretical predictions, with fidelities generally surpassing 90\%. These experimental results are very promising with regards to the potential future applications of the conversion gate, which includes converting between different small-sized multipartite entangled states. This is an important problem in quantum networks, where different quantum protocols require different resource states and the conversion gate can be used to prepare them.  The conversion gate can also be used to rewire the entanglement connections in larger multipartite entangled states in the form of extended graph states~\cite{Tashima} and so it would provide a useful way to reconfigure a network for a given distributed quantum protocol.

{\it Acknowledgments.---}
M.~M., R.~S., M.~Mikov\'{a}, I.~S. and  M.~J. acknowledge support from the Czech Science Foundation (GA16-17314S). P.~M. acknowledges support from the Czech Science Foundation (GB14-36681G). T.~T. and M.~T. acknowledge support from the South African National Research Foundation and the South African National Institute for Theoretical Physics.


\end{document}